\documentclass[twocolumn,showpacs,preprintnumbers,amsmath,amssymb]{revtex4}
\usepackage{graphicx}
\usepackage{dcolumn}
\usepackage{bm}

\begin{document}


\title{Local Entanglement and quantum phase transition in spin models}

\author{Shi-Jian Gu$^{1}$}\author{Guang-Shan Tian$^{1,2}$}
\author{Hai-Qing Lin$^{1}$}

\affiliation{$^1$Department of Physics and Institute of Theoretical Physics,
The Chinese University of Hong Kong, Hong Kong, China} \affiliation{$^2$School
of Physics, Peking University, Beijing 100871, China}

\date{July 31, 2005}

\begin{abstract}

Due to the phase interference of electromagnetic wave, one can recover the
total image of one object from a small piece of holograph, which records the
interference pattern of two laser light reflected from it. Similarly, the
quantum superposition principle allows us to derive the global phase diagram of
quantum spin models by investigating a proper local measurement. In the present
paper, we study the two-site entanglement in the antifferomagnetic spin models
with both spin-$1/2$ and $1$. We show that its behaviors reveal some important
information on the global properties and the quantum phase transition of these
systems.

\end{abstract}


\maketitle

\section{Introduction}

Quantum entanglement in the ground-state of strongly correlated systems
\cite{TJOsbornee,AOsterloh2002,XWang01,SJGu03,SJGu05,JVidal04,LFZhang04,REryigit04,FCAlcaraz04,XPeng05,LAWu04,MFYang05,ARIts05,RXin05,XFQian05,HFan04,SBLixxx,FVerstraete04,BQJin04,MPopp05,GVidal2003,VEKorepin04,YChen04,SJGurecent1,JWang03,AAnfossi05}
has been intensively studied in recent years. Its non-trivial behavior in these
system at quantum phase transition point \cite{Sachdev} attracted many
physicists' interest. Most of previous works focused on the spin models. And
the concurrence, a measure of entanglement of two qubits \cite{Hill}, has been
widely used in studying these systems. For example, Osterloh {\it et.al}
\cite{AOsterloh2002} studied the concurrence between two spins located on a
pair of nearest-neighbor sites in the transverse-field Ising model
\cite{TJOsbornee}. They found that this quantity shows singularity and obeys
the scaling law in the vicinity of the quantum phase transition point of the
system. On the other hand, for the XXZ model, the concurrence is a continuous
function of the anisotropic parameter and reaches its maximum at the critical
point \cite{SJGu03,SJGu05}.

However, the concurrence is very short-ranged. It vanishes quickly as the
distance between two sites increases. For a substitution, Verstraete {\it et
al} \cite{FVerstraete04} proposed the concept of localizable entanglement. They
showed that it is long-ranged in quantum spin systems
\cite{FVerstraete04,BQJin04,MPopp05}. Alternatively, studies on the block-block
entanglement between two parts of the system have established a connection
between conformal field theory and the critical phenomena in the condensed
matter physics \cite{GVidal2003,VEKorepin04}. Furthermore, since real systems
consist of itinerant electrons, some authors generalized this concept to the
entanglement in lattice fermion systems \cite{SJGurecent1,JWang03,AAnfossi05}.
For example, for the extended Hubbard model, we showed that its global phase
diagram can be identified by the local entanglement \cite{SJGurecent1}.
Therefore, one is convinced that the entanglement of the ground state, like the
conductivity in the Mott-insulator transition and quantum Hall effect, and
magnetization in the external-field-induced phase transition, plays also a
crucial role in quantum phase transitions.

In classical optics, by recording interference pattern of two laser reflected
from one object, its global information can be recovered from a piece of
holograph due to the classical interference. Similarly, the reduced density
matrix of a quantum system contains not only information on its subset, but
also the correlation between this subset and the rest part of the system. This
fact allows us to study some global properties of the system by investigating a
part of it, as we can see from the nontrivial behavior of block-block
entanglement \cite{GVidal2003,VEKorepin04} in the quantum spin models, as well
as the local entanglement in extended Hubbard model \cite{SJGurecent1}.
However, for the block-block entanglement, the size of the reduced density
matrix, which is required for calculation, grows exponentially as diameter of
the block increases. Obviously, it is impossible and, also, unnecessary to
consider the case in which the size of block is comparable to macroscopic
length. On the contrary, we would like to derive some properties of the system
in a simple way. Therefore, we shall consider the block which contains
sufficient information to reveal some global information of the system.
\emph{Local entanglement, as a limiting case of block-block entanglement, plays
such a role}.

The main purpose of this paper is to study the entanglement between a local
part and the rest part of the system in the ground-state of quantum spin models
for both spin $S=1/2$ and 1. Technically, to show the change of symmetry in the
ground state of the system, we need to investigate the entanglement between a
pair of neighboring spins and the rest part of the system. It is due to the
fact that the ground states of these models are usually spin singlet state. Our
intention is to show that the global properties of the system can be also well
understood from the behavior of local part due to the quantum coherence.
Although the local part consists only of two spins, we find that it is
sufficient to describe the quantum phase transition in these systems.
Furthermore, we show that the change of symmetry in the ground state at the
isotropic antiferromagnetic transition point yields a maximum value of local
entanglement. On the other hand, at the ferromagnetic transition point, the
singular behavior of the local entanglement can be clarified from the point
view of infinite degeneracy and level-crossing. In the two-dimensional case,
the cusp-like behavior of the local entanglement at the transition points
implies the existence of long-range correlation in the thermodynamic system,
which is absent in the one-dimensional case.

\section{One-dimensional spin-1/2 XXZ model}

First, we consider the spin-1/2 system. The corresponding Hamiltonian of the
antiferromagnetic XXZ model reads
\begin{equation}
\hat{H} = \sum_{\langle{\bf ij}\rangle} \left(\hat{S}_{\bf i}^x \hat{S}_{\bf
j}^x + \hat{S}_{\bf i}^y S_{\bf j}^y + \Delta \hat{S}_{\bf i}^z \hat{S}_{\bf
j}^z\right),\label{eq:Hamiltonian}
\end{equation}
where $\hat{S}_{\bf i}^x,\>\hat{S}_{\bf i}^y$ and $\hat{S}_{\bf i}^z$ are
spin-1/2 operators at site $\bf i$ and $\Delta=J_z/J_{x}\; (J_x=J_y)$ is a
dimensionless parameter characterizing the anisotropy of the model. The sum
is over all pairs of nearest-neighbor sites $\bf i$ and $\bf j$.
We impose the periodic boundary condition on the system. Thus, the choice
the neighboring spin pairs is independent of site index.

It is not difficult to prove that the Hamiltonian commutes with the
$z$-component of total spins $\hat{S}^z_{\rm total}=\sum_{\bf i}\hat{S}^z_{\bf
i}$. Therefore, the eigenstate of the Hamiltonian is also the eigenstate of the
$z$-component of total spin operator $\hat{S}^z_{\rm total}$.
As a result, the reduced density matrix of the neighboring spins
has the following form
\begin{eqnarray}
\hat{\rho}_{\bf \langle ij\rangle} = \left(
\begin{array}{llll}
u^+ & 0 & 0 & 0 \\
0 & w_1 & z & 0 \\
0 & z^* & w_2 & 0 \\
0 & 0 & 0 & u^-
\end{array}\label{eq:reduceddensitymatrix}
\right)
\end{eqnarray}
with respect to the standard basis $|\uparrow\uparrow\rangle,
|\uparrow\downarrow\rangle, |\downarrow\uparrow\rangle,
|\downarrow\downarrow\rangle$. Furthermore, the elements in the density matrix
$\rho_{\bf \langle ij\rangle}$ can be expressed in terms of the correlation
functions. We have
\begin{eqnarray}
&& u^\pm=\frac{1}{4}\pm \langle \hat{S}^z_{\bf i} \rangle +\langle
\hat{S}^z_{\bf i}\hat{S}^z_{\bf
j}\rangle, \nonumber \\
&& z=\langle \hat{S}^x_{\bf i} \hat{S}^x_{\bf j}\rangle + \langle
\hat{S}^y_{\bf i} \hat{S}^y_{\bf
j}\rangle, \nonumber\\
&& w_1=w_2=\frac{1}{4} -\langle \hat{S}^z_{\bf i} \hat{S}^z_{\bf
j}\rangle\
\label{eq:elementofmatrix}
\end{eqnarray}

It is well known that, when $\Delta>-1$, the global ground-state of the XXZ
model on the cubic lattice is nondegenerate and is a spin singlet
\cite{Lieb,Affleck}. It implies that $\langle \hat{S}^z_{\bf i} \rangle=0$ in
Eq. (\ref{eq:elementofmatrix}). Therefore, entanglement of the two spins and
the rest part of the system can be characterized by the von Neumann entropy of
the reduced density matrix (\ref{eq:reduceddensitymatrix}). We have
\begin{eqnarray}
E_{v}&=&-u^+\log_2 u^+ -u^-\log_2 u^- \nonumber \\ && -\lambda
^+\log_2\lambda^+ -\lambda^-\log_2\lambda^-,
\label{eq:entropy}
\end{eqnarray}
where $\lambda^\pm = w_1 \pm z $. Obviously, the local entanglement combines
all three correlation functions $\langle \hat{S}^x_{\bf i} \hat{S}^x_{\bf
j}\rangle$, $\langle \hat{S}^y_{\bf i} \hat{S}^y_{\bf j}\rangle$ and $\langle
\hat{S}^z_{\bf i} \hat{S}^z_{\bf j}\rangle$.
They behave differently as symmetry of the system changes.
In the following, we study how the local entanglement
is affected by them.

We start with the one-dimensional XXZ model, which can be solved
exactly by the quantum inverse scattering method\cite{VEKorepin93,HABethe31,MTakahashib}.
Its Bethe ansatz solution reads\cite{HABethe31}
\begin{eqnarray}
\left(\frac{\sinh\gamma(\lambda_j+i)}
           {\sinh\gamma(\lambda_j-i)}\right)^N
           =\prod_{l\neq j}^{M}
   \frac{\sinh\gamma(\lambda_j-\lambda_l + 2i)}
        {\sinh\gamma(\lambda_j-\lambda_l - 2i)}
\label{app_eq:BAE}
\end{eqnarray}
where the parameter $\gamma$ arises from the anisotropic scale $\Delta$, i.e.,
$\Delta=\cos 2\gamma$, and $\lambda_j (j=1,\cdots, M)$ are spin rapidities,
which describe the kinetic behavior of a state with $M$ down spins. The regime
$-1<\Delta<1$ is characterized by real positive $\gamma$ while the regime
$1<\Delta$ by pure imaginary $\gamma$ with positive imaginary part. From the
Bethe ansatz equation, the energy spectra and the thermodynamics of the system
can be obtained from the solution of $M$ spin rapidities\cite{MTakahashib}.
In particular, we find the ground-state energy as a function of $\Delta$.
Consequently, the correlation functions can be calculated
by the Hellman-Feynman theorem. It yields
\begin{eqnarray}
&&\langle \hat{S}^z_{\bf i}\hat{S}^z_{\bf i+1}\rangle=\frac{1}{N}\frac{\partial
E(\Delta)}{\partial \Delta}  \\ \nonumber &&\langle \hat{S}^x_{\bf
i}\hat{S}^x_{\bf i+1}\rangle =\langle \hat{S}^y_{\bf i}\hat{S}^y_{\bf
i+1}\rangle=\frac{1}{2}\left(\frac{E(\Delta)}{N}-\Delta \langle \hat{S}^z_{\bf
i}\hat{S}^z_{\bf i+1}\rangle\right).
\end{eqnarray}

\begin{figure}
\includegraphics[width=7cm]{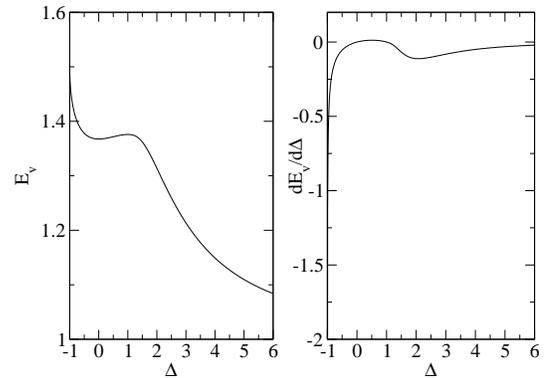}
\caption{\label{figure_lev1D} The local entanglement (left) of one-dimensional
XXZ model and its first derivative (right) as a function of the anisotropic
term $\Delta$.}
\end{figure}

\begin{figure}
\includegraphics[width=7cm]{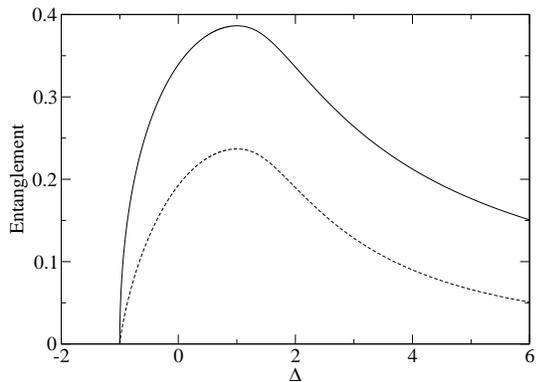}
\caption{\label{figure_entoff} The concurrence (solid line) and the
entanglement of formation (dashed line) between two nearest neighbors in
one-dimensional XXZ model as a function of the anisotropic term $\Delta$.}
\end{figure}

For one-dimensional spin-1/2 XXZ model, two critical points separate the whole
region of $\Delta$ into three different phases. These are the ferromagnetic
phase for $\Delta < -1$, the metallic phase for $-1 < \Delta < 1$ and the
antiferromagnetic insulating phase for $\Delta > 1$. In the ferromagnetic
region, the ground state is doubly degenerate and the local entanglement is
zero. It means that there is no quantum correlation between two parts of the
system. Therefore, we shall concentrate on the region of $\Delta>-1$. In the
Ising limit $\Delta\rightarrow\infty$, the ground state is approximately a
superposition of two N\'{e}el states. Therefore, the quantum fluctuation is
zero, i.e., $\langle \hat{S}^x_{\bf i} \hat{S}^x_{\bf i+1}\rangle=0$. Moreover,
the longitudinal antiferromagnetic correlation $\langle\hat{S}^z_{\bf i}
\hat{S}^z_{\bf i+1}\rangle=-1/4$. Consequently, the local entanglement tends to
unit, i.e. $E_v(\Delta\rightarrow\infty)=1$, in this limit. It is due to the
fact that, as $\Delta$ decreases, the effect of hoping process become more and
more important. Thus the fluctuation term $\langle \hat{S}^x_{\bf i}\hat{
S}^x_{\bf i+1}\rangle=0$ is undrawn from zero, while the ordering term is
suppressed. Therefore, the quantum correlation between the local part and the
rest part of the system will be enhanced. On the other hand, when $\Delta=0$,
the Hamiltonian can be transformed into a free spinless fermion model through
the Jordan-Wigner transformation. In this case, the nearest-neighbor
correlation functions of the system are $\langle S^x_{\bf i} S^x_{\bf
i+1}\rangle=\langle S^y_{\bf i} S^y_{\bf i+1}\rangle=-1/2\pi$ and $\langle
S^z_{\bf i} S^z_{\bf i+1}\rangle=-1/\pi^2$. Therefore, we have
$E_v(\Delta=0)=1.3675$. Obviously, the fluctuation term is dominant, as we can
see from the two eigenvalue of the reduced density matrix: $\lambda^+=0.669$
and $\lambda^-=0.033$. By Eq.~(\ref{eq:entropy}), it produces a relatively
smaller entropy. In this situation, both ferromagnetic and antiferromagnetic
couplings are expected to weaken the dominant $\lambda^+$. Therefore, when the
antiferromagnetic coupling is turned on, the competition between ordering and
fluctuation will lead to a maximum local entanglement at a certain point. As we
can see from Fig. \ref{figure_lev1D}, it is just the critical point $\Delta=1$
where the quantum phase transition undergoes. By the Bethe-ansatz
result\cite{MTakahashib}, the correlation functions at $\Delta=1$ take on
values $\langle \hat{S}^x_{\bf i} \hat{S}^x_{\bf i+1}\rangle=\langle
\hat{S}^y_{\bf i} \hat{S}^y_{\bf i+1}\rangle=\langle \hat{S}^z_{\bf i}
\hat{S}^z_{\bf i+1}\rangle=(1/4-\ln 2)/3$. Then, the local entanglement becomes
$E_v(\Delta=1)=1.3759$. Moreover, the local entanglement is a continuous
function around the transition point. As $\Delta$ becomes negative, the
antiferromagnetic order is obviously suppressed and the elements $u^\pm$ in
$\rho_{\bf ij}$ tends to $1/4$. This will also result in a larger value of
$E_v$. Another interesting result is the singular behavior of $E_v$ at
$\Delta=-1$, as shown in Fig.~\ref{figure_lev1D}. It is due to the
infinite-degeneracy at the critical point. Thus, in one-dimensional XXZ model,
the quantum phase transitions can be well described by the local entanglement.
However, these properties cannot be identified by the ground-state energy,
which is usually used in analyzing the quantum phase transitions. For instance,
at $\Delta=1$, the ground-state energy is actually a continuous function and
shows no sign of singularities \cite{CNYang66}.

Clearly, the concept of the local entanglements is quite different from the one
of the pairwise entanglement, usually measured by the concurrence\cite{Hill}.
They describe the quantum correlation between different part of the system.
However, they are closely to each other. To make this point more clear, we show
both the concurrence and the local entanglement of the present model as a
function of $\Delta$ in Fig.~\ref{figure_entoff}. In the region of
$0<\Delta<\infty$, as we explained above, the competition between ordering and
fluctuation leads to a maximal value for both the local entanglement and the
concurrence. As $\Delta\rightarrow -1^+$, however, the concurrence is dropped
down to zero and the local entanglement is pulled up dramatically. This fact is
consistent with the monogamy properties of the quantum
entanglement\cite{VCoffman2000}. On the other hand, the concurrence is only
valid for the description of entanglement between two qubits. Therefore, for
fermionic models and spin models with spin larger than 1, it is not a well
defined measurement for pairwise entanglement, although the
negativity\cite{GVidal2002} may partially characterize it to a certain extent.
In order to study the role of quantum correlation played in the quantum phase
transition, the local entanglement, as the simplest and computable measurement
of the block-block entanglement is a good replacement, as we showed for the
extended Hubbard model\cite{SJGurecent1}.

\section{Two- and three-dimensional spin-1/2 XXZ model}

\begin{figure}
\includegraphics[width=7cm]{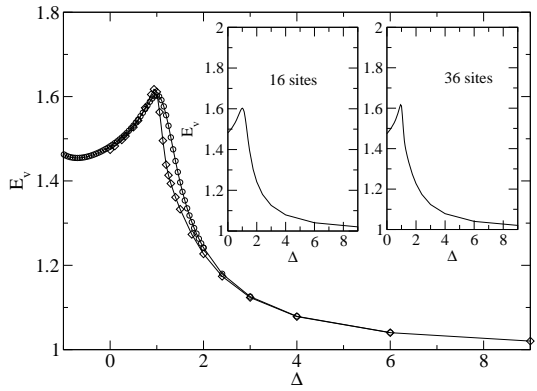}
\caption{\label{figure_len36} The local entanglement of two-dimensional XXZ
model as a function of the anisotropic term $\Delta$ for $4\times 4$ (circle
 line) and $6\times 6$ (square line) square lattice.}
\end{figure}

Next, we extend our study to two-dimensional antiferromagnetic spin-1/2 XXZ
model. Since there is no exact solution for higher dimensional XXZ model, some
approximate approaches such as the spin-wave theory\cite{JEHirsch89,WZheng91}
or numerical calculations on finite lattice\cite{hql-ed}, have to be applied.
In fact, with the help of scaling analysis, the results of the exact
diagonalization are fully consistent with those obtained by the quantum Monte
Carlo\cite{Sandvik} and other analytical method\cite{JEHirsch89,WZheng91}. By
the exact diagonalization technique, we calculate the local entanglement of the
two-dimensional model as a function of $\Delta$ on both $4\times 4$ and
$6\times 6$ square lattices with periodic boundary condition and show the
results in Fig.~\ref{figure_len36}. For $\Delta>1$, the local entanglement is a
decreasing function of $\Delta$. It is due to the fact that enhancement of the
antiferromagnetic order suppresses the quantum correlation between a local part
and the rest part of the system. On the other hand, for $0<\Delta <1$, it is an
increasing function of $\Delta$, as we explained above for the one-dimensional
case. Furthermore, the local entanglement reaches its maximum at the isotropic
point $\Delta=1$. However, unlike its one-dimensional counterpart, the local
entanglement of the two-dimensional XXZ model shows a cusp-like behavior around
the critical point. Such singular behavior implies the existence of
long-range-order, which is absent in the one-dimensional case. It is consistent
with our previous result for the two-dimensional XXZ model\cite{SJGu05}.

\begin{figure}
\includegraphics[width=7cm]{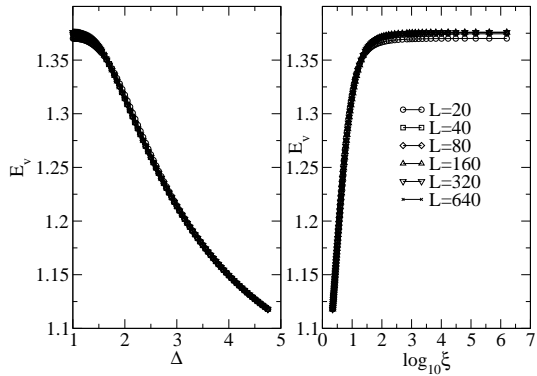}
\caption{\label{figure_slent} The representation of $E_v$ as a function of
$\Delta$ (left) and correlation length \cite{RJBaxterb,SJGu2002}($\xi$ is in
unit of lattice constant) for one-dimensional XXZ model in the insulating phase
($\Delta > 1$) with different size.}
\end{figure}

As is well known, the ground state of antiferromagnetic XXZ model is
nondegenerate. One of our main results on the one-dimensional XXZ model is that
the local entanglement is a smooth continuous function of $\Delta$ around the
critical point. In the meantime, the correlation functions decays by power-law
in the antiferromagnetic region. Therefore, the local quantities of the system,
such as energy density and the nearest-neighbor correlation functions, are not
effected by the those spins, which are far away from the local pair. As a
result, the finite degree of freedom in small system guarantees the analyticity
of these quantities \cite{CNYang66}. In particular, the concurrence is
continuous \cite{SJGu03}. For the same reason, the local entanglement should be
also continuous. Furthermore, in Fig. \ref{figure_slent}, we show the scaling
behavior of the local entanglement for $\Delta >1$. We find that the local
entanglement does not obey a length scaling law. It is rather different from
behavior of other quantities, such as the spin stiffness in the XXZ model
\cite{SJGu2002}. However, in two-dimensional XXZ model, the maximum of local
entanglement at the critical point is sharpened, as the size of the system
increases (See Fig. \ref{figure_len36}). This is due to the existence of
long-range correlation, which leads to a strong dependence of local
entanglement on the system size. In other words, it is the infinitely large
degree of freedoms in the infinite system which introduce the singular behavior
of the local entanglement. These results are also consistent with the previous
conclusions on the concurrence of the two-dimensional XXZ model
\cite{SJGu05,Sandvik,JEHirsch89}.

\section{One dimensional spin-1 model}

\begin{figure}
\includegraphics[width=7cm]{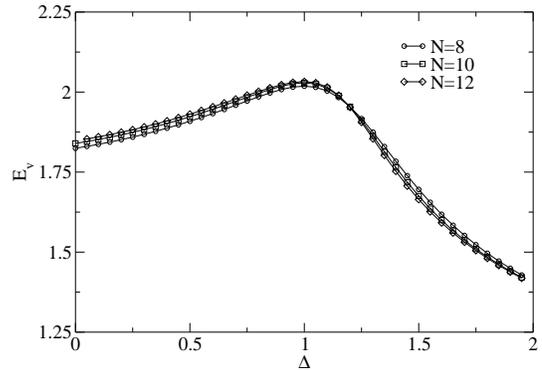}
\caption{\label{figure_leone} The local entanglement of one-dimensional XXZ
model as a function of the anisotropic term $\Delta$ for various system size
$L=8, 10, 12$.}
\end{figure}

\begin{figure}
\includegraphics[width=7cm]{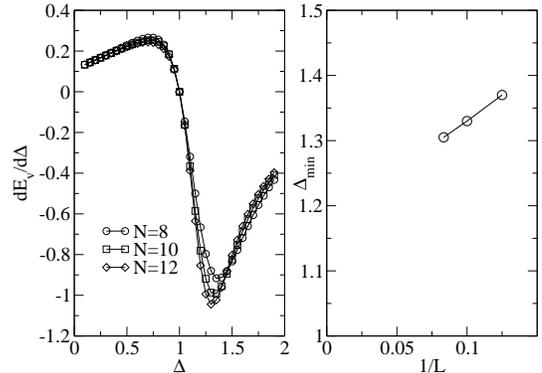}
\caption{\label{figure_dleone} LEFT: The first derivative of the local
entanglement of one-dimensional XXZ model as a function of the anisotropic term
$\Delta$ for various system size $L=8, 10, 12$; RIGHT: The scale analysis of
the minimum point of its first derivative.}
\end{figure}

\begin{figure}
\includegraphics[width=7cm]{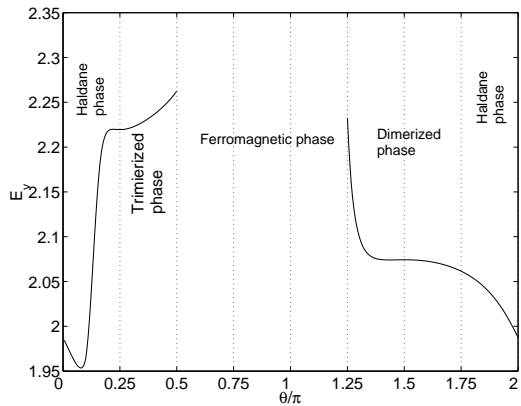}
\caption{\label{biqu6_entang} The local entanglement of one-dimensional
bilinear-biquadratic model as a function of the $\theta$ for system size
$L=6$.}
\end{figure}

In this section, we extend our study to a one-dimensional spin-1 model,
whose Hamiltonian is defined as
\begin{equation}
\hat{H} = \sum_{\langle{\bf ij}\rangle} \left(\hat{S}_{\bf i}^x
\hat{S}_{\bf j}^x + \hat{S}_{\bf i}^y S_{\bf j}^y + \Delta \hat{S}_{\bf
i}^z \hat{S}_{\bf j}^z - \beta \left( \hat{S}_{\bf i} \cdot \hat{S}_{\bf j}
\right)^2 \right),
\label{eq:Hamiltonian_2}
\end{equation}
where $\beta$ is a real parameter. Differing from the case of spin-1/2,
the ground-state of spin-1 chain contains fascinating surprise
which originates from the Haldane gap.
For the isotropic antiferromagnetic Heisenberg chain with integer
spin, the existence of this gap in the excitation spectrum
of the system was first conjectured by Haldane \cite{FDMHaldane83}.
Then, it was confirmed by the quantum Monte Carlo simulation\cite{MTakahashi89}
and scaling analysis from the exact diagonalization\cite{Spinoneexact}.
However, as $\Delta$ increases, this gap will vanishes around $\Delta\simeq
1.18$.

To study the role of the local entanglement in connection to the critical
behavior of this system, we use the exact diagonalization method to compute the
two-site entanglement in its ground state. The results are shown in
Fig.~\ref{figure_leone}. Obviously, the behavior of the local entanglement is
quite similar to the one of spin-1/2 model. In particular, the local
entanglement has a maximum at the isotropic point, although it is not a quantum
transition point. A careful scrutiny reveals that the transition occurring in
spin-1 chain is of the Kosterlitz-Thouless type, which is quite different from
the one for the Heisenberg model of spin-$1/2$. In the latter case, the spin
excitation spectrum of the system is gapless on both sides of the transition
point. However, for the system with spin-$1$, the spectrum becomes gapful on
one side of the transition point. This situation is very similar to the one
observed for the transverse field Ising model\cite{Sachdev}. Since the first
derivative of the concurrence is singular and obeys scaling law around the
transition point of this model\cite{AOsterloh2002}, we speculate that the same
behavior may be seen for the local entanglement around the transition point for
the spin-$1$ Heisenberg chain. Indeed, by taking the first derivative of the
local entanglement of Hamiltonian (\ref{eq:Hamiltonian_2}) with respect to
$\Delta$, we find a minimum point around $\Delta\approx 1.3$ in
Fig.~\ref{figure_dleone}. Moreover, its value varies as the size of the system
increases. By fitting data with respect to the sample size, we see that it
tends to $\Delta\simeq 1.18$ in the thermodynamic limit.

To explore the effect of parameter $\beta$,
we set $\Delta=1$ and rewrite Hamiltonian (\ref{eq:Hamiltonian_2}) as
\begin{equation}
\hat{H} = \sum_{\langle{\bf ij}\rangle}
\left(\cos\theta\hat{S}_{\bf i} \cdot \hat{S}_{\bf j}
+ \sin\theta\left(\hat{S}_{\bf i} \cdot
\hat{S}_{\bf j} \right)^2\right).
\label{eq:Hamiltonian_3}
\end{equation}
By introducing trigonometric functions, we are also able to take the effect of
coupling sign into our consideration. This Hamiltonian has a rich phase diagram
at zero temperature. It consists of the Haldane phase, trimerized phase, and
dimerized phase\cite{KNOmura91}. We now study the role of local entanglement
around critical points.

Our results are shown in Fig. \ref{biqu6_entang}. From the figure, we find that
the local entanglement reaches a local minimum at $\theta=\pi/4$, which
separates the Haldane phase and trimerized phase. At $\pi/2<\theta<5\pi/4$, the
ground state is ferromagnetic and degenerate. In this case, the local
entanglement cannot be well defined because the thermal ground state comprises
all states of lowest energy with equal weight. Indeed, a fully polarized state
is separable. Therefore, if the ground state is spin polarized one, its
entanglement is zero. As a result, sudden changes in the local entanglement
occur at both $\theta=\pi/2$ and $\theta=5\pi/4$. They are caused by the
ground-state level-crossing. However, we do not find any discernible structure
around the critical point $\theta=7\pi/4$, which separates the dimerized phase
from the Haldane phase. Since we consider only the sample with $L=6n$ sites,
whose ground state has both of trimerized and dimerized orderings, the rapid
increasing size imposes further limit on scaling analysis. On the other hand,
we expect that an extremum will appear in its derivatives around
$\theta=7\pi/4$ when the system size becomes large, just like the case of
$\Delta\simeq 1.18$ in XXZ model. We also find a local minimum at
$\theta=3\pi/2$. Around this point, $\cos\theta=0$ and the Hamiltonian is
reduced to $\hat{H} = -\sum_{\langle{\bf ij}\rangle} \left(\hat{S}_{\bf i}
\cdot\hat{S}_{\bf j} \right)^2$. Its ground state is nondegenerate, while its
first excited state is $3$-fold degenerate on one side of the transition point
and $5$-fold on the other side, the degeneracy is exactly $8$-fold at the
transition point. Moreover, in the previous works \cite{KNOmura91}, it has been
shown that the dimerized phase in the region $5\pi/4<\theta <3\pi/2$ is
gapless. Therefore, a level-crossing between the lowest excited states must
occur, as the one observed for the XXZ model of spin-$1/2$ at the transition
point $\Delta=1$. Since the minimum in local entanglement is intrinsically
related to the symmetry of the Hamiltonian, we believe that the point
$\theta=3\pi/2$ is also a critical point at which two ordered phases are
separated.

\section{Summary and Acknowledgment}

In the present paper, we study the global phase diagram of the quantum spin
models with either spin-$1/2$ or $1$ by investigating the local entanglement.
We show that, indeed, many global properties of the system can be derived from
such a local measurement. In fact, one has observed a long time ago that the
original three-dimensional image can be recovered from a small piece of
holograph, though its resolution is reduced. It is caused by the classical
interference. Similarly, our findings can be understood on the basis of the
quantum superposition principle.

We see that, for one-dimensional spin-1/2 XXZ model, the local entanglement
shows singular behavior around one critical point $\Delta=-1$ and takes on its
maximum at another critical point $\Delta=1$. For the two-dimensional spin-1/2
XXZ model, we find that the maximal point of local entanglement around
$\Delta=1$ is sharpened. It is due to the existence of long-range order, which
is absent in one-dimensional case. For the spin-1 XXZ system, the local
entanglement also has maximum at the isotropic point. Moreover, the scaling
analysis manifests that its first derivative will tend to the critical point
$\Delta\approx 1.18$ as the system size becomes infinite. The rich phase
diagram for the bilinear-biquadratic model can be almost mapped out from the
behavior of the local entanglement. Furthermore, from the singular behavior the
local entanglement, we find that the point $\theta=3\pi/2$ may be also a
transition point of this model. This issue deserves definitely further
investigation.

This work was supported by a grant from the Research Grants Council of the
HKSAR, China (Project No. 401703) and the Chinese National Science
Foundation under Grant No. 90403003.

\end{document}